\documentstyle[preprint,aps]{revtex} 
\def\Bi{$Bi_2Sr_2CaCu_2O_{8+y} $}
\begin{document}
\draft
\widetext
\title{ The extended Hubbard model applied to phase diagram  and
the pressure effects in \Bi superconductors     }
\author{E. V.L. de Mello\cite{email} } 
\address{Departamento de F\'isica, Universidade Federal Fluminense, 
av. Litor\^ania s/n, Niter\'oi, R.J., 24210-340, Brazil}
\date{\today}
\maketitle
\begin{abstract}
We  use  the two dimensional extended Hubbard
Hamiltonian  with the position of the attractive potential as a
variable parameter  with  a  BCS type approach  to study  the interplay 
between the superconductor transition  temperature $T_c$ and  hole 
content for  high temperature superconductors. 
This novel method gives some insight on the range and intensity 
of the Cooper pair interaction and why different compounds have different
values for  their  measured coherence lengths and
it describes well the experimental results of the superconducting
phase diagram $T_c \times n$.
The calculations  may also be used to study the effect of the applied pressure
with the assumption that it increases the attractive potential which is
accompanied by an  increase in the superconductor gap.
In this way we obtain a microscopic interpretation for the
intrinsic term and a general expansion for $T_c$ in terms of the pressure
which reproduces well the experimental measurements on the \Bi superconductors.
\end{abstract}
\pacs{71.10.Fd,74.62.Fj,74.25.Dw}
\vspace{2cm}
{\bf Key Words}: BCS Model, Hubbard Hamiltonian, Phase Diagram, Pressure Effects\\
\newpage



\section{Introduction}

 In order to understand the superconductivity mechanism for high-$T_c$ 
superconductors (HTSC), many experiments have attempted to find out
possibles correlations between the superconductivity temperatures $T_c$ 
and some physical parameter. Some of this experiments deal with 
the interplay among $T_c$, hole contents and pressure effects 
which have revealed interesting features like the maximum attained 
$T_c$\cite{Gao}
and have attracted a lot of attention as it is summarized by 
some  review articles\cite{Griessen,Taka,Schill}. On the other hand, neither
one of the several microscopic theoretical models have been able to
deal successfully with all these observed basic  properties and therefore 
the nature of the condensate pair has not yet been established. As
a consequence, some phenomenological approaches  have been advanced in an
effort to account for the large experimental data on HTSC. For instance,
the $T_c$  dependence  on the number of carriers per $CuO_2$ 
follows the so-called universal inverse parabolic curve, i.e.,
$T_{c}=T_c^{max}[1-\eta(n-n_{op})^{2}]$
where $n_{op}$ is the optimum carrier density $n$ and $\eta$ is an adjustable
parameter which depends on the type of compound. 

 Recently we have introduced an intermediate approach\cite{Mello1} which 
is neither pure microscopic nor pure phenomenological: it starts with  the
microscopic extended Hubbard Hamiltonian 
(t-U-V) on a square lattice of side $a$ 

\begin{equation}
H=-\sum_{\langle ij \rangle , \sigma}t(c^{\dagger}_{i \sigma} c_{j \sigma}
 + h.c.)+U \sum_{i} n_{i \downarrow}n_{i \uparrow}-V \sum_{\langle ij
 \rangle}n_{i}n_{j} .
\label{Hamiltonian}
\end{equation}
where $t$ is the nearest neighbor hopping integral, U is the on-site 
repulsion and V  a phenomenological attractive interaction. Several authors 
have argued that this Hamiltonian  accounts  for the physics of the 
charge carriers\cite{MRR,Angilella} in HTSC. The strength of the 
phenomenological potential coupling 
constants are determined {\it a posteriori} in comparison with the 
experimental results for $T_c \times n$. The hopping integral 
$t$ is obtained  by comparing with band structure calculations\cite{Angilella}. 
Using the BCS method and
the Fourier transform of the potential we have derived  zero and
finite temperature gap equations which are integrated up to
region where pair formation has a non-negligible probability. In
this way we have been able to relate $T_c$, hole content $n$  and
the strength of the zero temperature gap $\Delta (0)$ as described
in detail in Ref.\cite{Mello1}. As concerns the phase diagram, 
its optimal doping   is very sensitive if we move V from its original nearest 
neighbor position for the $s_{x^2+y^2}$-channel but does not change its 
optimal position for $d_{x^2-y^2}$-wave. A study of these and others channels 
as well as their complex mixtures and interplay as function of the position 
of V will be published elsewhere\cite{Mello15}.
This method was successfully applied to the
experimental measurements of the curves $T_c \times n$ for the
$La(Sr,Ba)CuO$ and $YBaCuO$ compounds\cite{Mello1}. On the other hand if 
we study the typical  parameter which characterizes the strength of the
attractive interaction for a square well in 2D with range $d$ and 
strength\cite{Landau} V, $\chi  = m_ed^2V/2\hbar^2$, we see that for the 
above two compounds and Hg1201 and \Bi,
it varies between 10-3000 and such high values indicate that  the size
of the bound states (or Cooper pairs) is related to the
position of the minimum and the range of
the attractive potential. This is the main motivation to change the
position of the attractive potential V from the usual nearest neighbor
and it is very interesting to find out that the curves for $T_c \times n$ 
yield a different $s_{x^2+y^2}$-wave optimal doping value depending on the
size of V  while it remains fixed for $d_{x^2-y^2}$-wave. 
Another interesting point from the theoretical point of view is that it provides
a natural interpretation for the order parameter expansion introduced by the
spin-fluctuation theory\cite{Pines}. In this
approach  it has been proposed a $d_{x^2-y^2}$-wave gap of the
form $\Delta_0(\vec k)(cosk_xa-cosk_ya)$ where $\Delta_0
(\vec k)$ is expanded in powers of $(cosk_xa+cosk_ya)$. Analising 
this expansion, one can easily verify that it contains
terms like $(cosk_xa-cosk_ya)(cosk_xa+cosk_ya)$ which are
proportional to  $(cos(2k_xa)-cos(2k_ya))$ and what can be seen as
a type of d-wave gap symmetry that arises from a potential
like $V(\vec k)=V_0(cos(2k_xa)+cos(2k_ya))$. By the same token, we
can find  terms proportional to  $(cos(3k_xa)-cos(3k_ya))$
which can be originated by a potential  $V(\vec k)=V_0(cos(3k_xa)+cos(3k_ya))$
and so on.  Based on this result, we adopt below a notation 
$V(\vec k)=V_0(cos(bk_xa)+cos(bk_ya))$, with $b=1,2,3,...$. As one expects, 
the more distant is the position of V, that is the larger is $b$, smaller is 
the value of the density necessary to start the
superconductivity process of pair formation.
In this way we have derived $T_c \times n$ curves\cite{Mello1} that agree well 
with the experimental data when V was placed at $6^{th}$ (s-wave)
and $3^{rd}$ (d-wave) 
neighbors for $La(Sr,Ba)CuO$  and $YBaCuO$ compounds  respectively 
which by the way is very close the ratios of their measured coherence 
length $\xi$\cite{Cyrot}. Thus the method provides a very interesting  and 
direct interpretation for the measured
values of the coherence length for the HTSC compounds besides the fact that it
gives a hint for the superconducting fundamental interaction in HTSC. 

We have also extended  this approach to study   the pressure 
effects\cite{Mello2} on HTSC. We
have noticed that $T_c(n)$ is very sensitive  to changes in 
the zero temperature gap amplitude  $\Delta (0)$
which is used as an adjustable parameter.
Consequently we have introduced the assumption that the pressure induces 
changes on the attractive potential end ultimately
on $\Delta (0)$\cite{Mello2}. This assumption together with the 
well documented fact  that the applied  pressure induces a charge transfer
from the reservoir layers  to the $CuO_2$ planes, has enabled us  
to derive   an expansion for $T_c(n,P)$ in power of the pressure. This
expansion was applied to the Hg1201, 1212 and 1232  compounds and has 
reproduced the measured  values of $T_c$ for a large set of hole content and
applied pressures up to 45 GPa\cite{Mello2}.
We should stress that since the principal interaction mechanism 
which forms Cooper pairs  has yet to be identified, the
connection between $T_c$, hole content, pressure effects
and a microscopic Hamiltonian should
provide  some clues for this interaction and it is a efficient  method to
obtain information on the microscopic mechanism for HTSC. Thus our study
provides a guide to the strength and range of this fundamental mechanism. 

 In this paper we shall clarify and illustrate  the main physical ideas of the 
method mentioned above  and to show that it is general and can be applied
to other HTSC than the  Hg-based compounds, as we  apply it to the experimental 
phase diagram and pressure data on Bi2212.
The  $Bi_2Sr_2CaCu_2O_{8+y}$ family
has been studied by several groups and in particular,  we shall discuss firstly
the measurements of the dependence of the critical temperature $T_c$ on
the hole content $n$\cite{Allgeier}  to set the pertinent parameters to be
used afterwards on the study of  the variation of $T_c$
under applied external pressures\cite{Huang}. The \Bi  is strongly
anisotropic, its resistivity on the ab plane is $10^4$ larger
than its similar along the c-axis\cite{Huang} which justify the use of a
2D model and has one of the shortest
coherence length\cite{Cyrot} among all cuprates superconductors and this
property will be taken into account to set the parameter $b$ in our 
calculations.

\section{The Method} 

Following the BCS method\cite{de Gennes} we use a many-body wave function 
which describes the formation of  pairs of charge carriers

\begin{equation}
\vert \Phi\rangle=\prod_{\vec k}\left(u_{\vec k}+v_{\vec k} a^{\dagger}
_{\vec k \uparrow} a_{-\vec k \downarrow}\right)\vert \Phi_0 \rangle,
\end{equation}
where $\vert \Phi_0\rangle$ is the empty band state and $u^{2}_{
\vec k}+v^{2}_{\vec k}=1$.

In connection with the Hamiltonian of Eq.\ (\ref{Hamiltonian}), and
the variational approach\cite{de Gennes}, we  obtain

\begin{equation}
\Delta_{\vec k}=-\sum_{\vec l}V_{\vec k\vec l}u_{\vec l}v_{\vec l}
\label{gapuv}
\end{equation}
where $V_{\vec k\vec l}$ is
the Fourier transform of the potential of Eq.\ (\ref{Hamiltonian}), 
and  following along the lines of Ref.\cite{Angilella}  
may be written in a  "separable" form,
$V_{\vec k\vec l}=U-2V\left(\cos(k_xa)\cos(l_xa)+\cos(k_ya)\cos(l_ya)\right)$.
A similar approach which takes next-nearest neighbor hopping into 
account was developed by Angilella et al\cite{Angilella} with 
the same purpose to study pressure effects on \Bi  compounds but
they only deal with nearest neighbor attractions.
With the use of the expressions for $u_{\vec l}$ and $v_{\vec k}$,
we get an  equation for the zero temperature gap,

\begin{equation}
\Delta_{\vec k}=-\sum_{\vec l}V_{\vec k\vec l}\frac{\Delta_{\vec l}}
{2\left(\xi^2_{\vec l}+\Delta^2_{\vec l}\right)^{1/2}} .
\label{gap}
\end{equation}
Where $\xi_{\vec k}=-2t(cos(k_{x}a)+cos(k_{y}a))-\mu$ , 

Thus since the gap has the same functional form of the potential, namely, 
$\Delta_{\vec k}= \Delta (0) (cos(k_{x}a)\pm cos(k_{y}a))/2$, where 
the plus sign is for the s-wave and the minus sign is for d-wave 
channel. The chemical potential $\mu$ yields the density $n$ that 
 must be calculated 
self-consistently\cite{MRR} but as it concerns the superconducting phase 
boundary it suffices to approximate it by the $T=0$K  value of 
the maximum energy (concentration dependent) in the tight-binding  band. 

In order to study pair formation, we calculate
the probability of finding
a hole pair, that is, the condensation amplitude $F_{\vec k}\equiv
u_{\vec k}v_{\vec k}$. It is a general result\cite{de Gennes} that
it has a maximum at $k_F$
and drops very rapidly for $\left\vert k\right\vert >k_F$.
According to these considerations and taking $\vec k =(0,0)$ in Eq.\
(\ref{gap}), we obtain for s-wave

\begin{eqnarray}
  A= \int_{0}^{\alpha_\Delta}\int_{0}^{\beta_\Delta}d\alpha d\beta 
  \frac{\left(U-2Vf_{+}(b\alpha,b\beta)\right)
   f_{+}(b\alpha,b\beta)/2}{\left(\left(2t\left(f_{+}(\alpha,\beta)
  -\mu \right)\right)^2+ 
    (\Delta(0)f_{+}(b\alpha,b\beta)/2)^2 \right)^{1/2}}\;  ,
\label{gap0}
\end{eqnarray}
where $f_{\pm}(\alpha,\beta)\equiv \cos\alpha \pm \cos\beta$ and
$\alpha=k_{x}a$ and $\beta=k_{y}a$. $\mu=f_{+}(\alpha_F,\beta_F)$ 
and $\alpha_F$ and $\beta_F$ are the
maximum $T=0$ occupied values (like a Fermi momentum) that
depend on the density of holes $n$. The parameter $b$ is 
used to set the position of the attractive potential V, thus {\bf  $b=2$ for
third nearest neighbor and so on}.  We use 
$\alpha_{F}= \beta_{F}\approx arcos(1-n)$ and $\mu=-2t(cos(\alpha_{F} ))$.
The integrations are
performed up  to $\alpha_\Delta$ and $\beta_\Delta$, which are
determined  at values where the condensation amplitude becomes very
small,  namely, $F_{\vec k} \approx 0.01$. This is
usually attained for $\xi_{\vec k}> 6\Delta(0)$.

For $T\not=0$, the
excitations with their respective probability must be taken into
account. The derivation of a self-consistently temperature-dependent
gap equation is analogous to that which leads to Eq.\ (\ref{gap0}).
At this point, we again follow the
BCS approach\cite{de Gennes} and assume that $\Delta(T)$ vanishes
at the critical temperature $T_c$, which yields the following equation
for s-wave symmetry,

\begin{eqnarray}
B= \int_{0}^{\alpha_\Delta}\int_{0}^{\beta_\Delta} d\alpha d\beta
         \frac{\left(U-2Vf_{+}(b\alpha,b\beta)\right) f_+(b\alpha,b\beta)\tanh
   \left(\frac{2\xi_{\alpha\beta}}{2K_{B}T_{c}}\right)}
  {2t\left(f_+(\alpha,\beta)-f_+(\alpha_F,\beta_F)\right)}\; ,
\label{gapT}
\end{eqnarray}
where we again integrate up to $\alpha_\Delta$ and $\beta_\Delta$

Now, if we had integrated Eqs.\ (\ref{gap0} and \ref{gapT}) over the whole
Brillouin Zone and solve them self-consistently, we would get $A=B=-(2\pi)^2$.
Instead of this procedure, we find, for a given density $n$, the value of $T_c$ 
which is compatible, i.e., make the equality $A=B$ for a 
given $\Delta(0)$ and  $b$. This is the basic procedure of our method.
The same approach can be carried out for d-wave symmetry but in this case
we take $\vec{k}=(0,\pi/2)$ in Eqs.(\ref{gap0} and \ref{gapT}). 
Thus, with only one given value for the parameter $b$ and 
another one for  $\Delta(0)$,
we can find the value of $T_c$ for a given density of carriers $n$ and 
therefore obtain  the curve $T_c \times n$.
 
\section{Pressure Effects}

Now we turn our attention to the study of the HTSC under pressure.
One of the effects of the  pressure which is generally
accepted and well documented in certain materials
is an increase of the carrier concentration on the $CuO_2$
planes transferred from the reservoir layers\cite{Griessen,Taka,Schill}.
Such pressure induced charge transfer (PICT) has been confirmed by 
Hall effect and
thermoelectric power measurements on several compounds\cite{Taka}.
Therefore this effect combined with an assumption of an  intrinsic
variation  of $T_c$
(linear on the pressure) independently of the charge
transfer was largely explored to account for the
quantitative relation between $T_c$ and the pressure P and it gave
origin to many models\cite{Griessen2,Almasan,Neumeier,Gupta,Mori}.
However, the various pressure data taken with 
the mercury family of compounds\cite{Cao,Gao} for underdoped and
overdoped compounds of $HgBa_2CuO_{4-\delta}$ (Hg1201) has some features 
that could not be interpreted\cite{Cao,Gao} by these pure charge transfer 
models by a single set of parameters. On the other hand we could
explain the Hg-family data using a single set of parameter and
this is an achievement of our method described below.

The  study of $T_c\times n$ which is necessary to find the appropriated values
of $b$ and $\Delta(0)$ (at $P=0$) demonstrated  the direct
proportionality  between $\Delta(0)$ and  $T_c$ which varies with the hole 
content $n$ and it is maximum at the optimal doping. Taking this result
into account and  the discussion of the above paragraph, 
we are led to propose that the effects of pressure are two-fold: 
(i)- The well accepted PICT and an additional assumption;
(ii)- A change in the attractive potential V as proposed previously by 
Angilella et al\cite{Angilella} which also
implies on a change in the zero temperature gap $\Delta(0)$. The assumption
"ii"  came from  the analysis of the curves $T_c \times n$
which have revealed that $T_c$ is proportional to $\Delta(0)$
but with a large constant of proportionality around the optimal 
doping\cite{Mello2} $n_{op}$ and smaller near the both
underdoped and overdoped extremes of the $T_c \times n$ curve.
The quantitative changes on V due to the applied pressure
were estimated  in Ref.\cite{Angilella} for the \Bi using the measured
values\cite{Huang} of the compressibility tensor under the 
linear approximation and the self-consistently equations. 

Thus the PICT (i) implies that $n(P)=n+\Delta n(P)$ and the assumption
of a pressure dependent gap (ii) implies that $\Delta(0,P)=\Delta(0)+
\Delta \Delta(0,P)$ and both equations lead to  
$T_{c}(n,P)=T_{c}(n(P),\Delta(0,P))$.
Therefore to estimate $T_c$ for a given compound  with a nominal  
value of $n$ and under a given pressure P,  we  perform an expansion of 
$T_c(n,P)$ in terms of P. With the assumption of the linear variation 
of $n$ and $T_c^{max}$ (or $\Delta(0)$) on the pressure,  we obtain only 
terms up to third order, that is,
\begin{mathletters}
\label{allequations}
\begin{equation}
T_c(n,P)=\sum_{Z=0}^3 \alpha_Z P^Z/Z!
\label{expP}
\end{equation}
with
\begin{equation}
\alpha_Z=({\partial \over \partial \Delta(0)}{ \partial \Delta(0)\over
 \partial P} + {\partial \over \partial n}{ \partial n \over \partial P})^Z
 T_c(n(P),\Delta(0,P))\:\; , z=1,2 \hspace{0.3cm} and \hspace{0.3cm} 3
\label{coeff}
\end{equation} 
\end{mathletters}
and $\alpha_0==T_{c}(n(0),\Delta(0))$ is just the $P=0$ term and 
where the derivatives $ \partial \Delta(0) \over
\partial P$ and $ \partial n \over \partial P$ are determined by 
comparing with the experimental data; $ \partial \Delta(0) \over
\partial P$ can be estimated by a set of values  $T_v \times P$ for a given
density $n$ and $ \partial n \over \partial P$ can be determined by Hall
coefficients or thermoelectric power measurements.  After determined these
parameters, one can derive analytical expressions for
each coefficient as function of $n$;  using the universal 
parabolic fitting  we can  obtain a analytical expression for 
$\partial T_c /\partial n$ and studying the phase diagram for different  
values of $\Delta (0)$, we can calculate $\partial T_c/\partial P$
which calculates  the $T_c$ dependency on P (assumption ii).
This procedure gives an intrinsic term which has a clear interpretation
since it comes from the changes in $\Delta(0)$
as well as a new third order term. It is remarkable that with a single set of
parameters calculated for HG1201,  with just the change in $T_c^{max}$,
we were also able to  apply the same expansion (with same coefficients)
to HG1212 and HG1223. 

\section{Comparison with the Experiments}

In order to apply the entire approach described above, we
start with the $P=0$ curve $T_c \times n$ for \Bi which allows us to
determine the initial parameters $b$ and $\Delta(0)$. We have used the
values $t=0.05 eV$ and V$=0.052 eV$ given in Ref.\cite{Angilella}. The coherence
length estimate for \Bi is $\approx 13\AA$ which is about 10\%
less than the $\xi$ for YBaCuO\cite{Cyrot}. Since we
have used  $b=2-3$ for fitting the phase diagram for these compounds,
we use here $b=2.6-2.7$ in order to obtain $T_c \times n$ for \Bi.
The results  are displayed on Fig.1 and one can see that it provides a very
reasonable fitting for the experimental points of Allgeier et al\cite{Allgeier}.
In order to match the $T_c^{max}=92.5$ with  V$=0.052 eV$, we used
$\Delta(0)=214$K which is the same for all values of $n$. This value is very
close $\Delta(0)=210$K used to the Hg1210  compounds because
both compounds have very closed values for $T_c^{max}=92-94$K and we used for
both a $s_{x^2+y^2}$ gap function.  We have
also plotted a parametric curve which has been applied to fit the experimental
points,  $T_{c}=T_c^{max}[1-\eta(n-n_{op})^{2}]$ with $T_c^{max}=92.5$K
and $\eta=9.0$. However the data of Huang et al\cite{Huang}
appears to have smaller
$T_c^{max}$,  displaced toward higher hole densities and they
are fitted with $\eta=25.0$. The parabolic is a pure phenomenological fitting 
which is used here only because it provides a analytical expression that
can be used in the pressure expansion to estimate  $\partial T_c/\partial n$.

Now that {\it  $b$ and $\Delta(0)$ has been determined} let us study the 
\Bi compound    under applied pressure.
From the analysis of the phase diagram\cite{Allgeier} shown in Fig.1 
and by varying the value of $\Delta(0)$ around the $P=0$ value of
$\Delta(0)=214$K, we can infer that $\partial T_c/\partial
\Delta(0)=0.35$ at $n\approx 0.25$ which is near the optimal doping 
value of hole content used by Angilella et al\cite{Angilella}. From
the Hall coefficient measurements\cite{Huang},
we  get $d(ln(n))/dP=+8\%$ and since their results seens to be
shifted to larger values of hole contents,   we estimate
$dn/dP\approx 3.0\times 10^{-2}$ {\it which is two order of magnitude higher 
than previously derived  $dn/dP\approx 1.8\times 10^{-3}$ for the pressure 
effects on Hg compounds\cite{Mello2}.}
From the  low pressure data we can get $\alpha_1=6.0K/GPa$
which allows us to derive the value $\partial \Delta(0)/\partial P=4.86
K/GPa$ which is necessary to determine all
the high order coefficients. We noticed that  this value is very close to 
$\partial \Delta(0)/\partial P=4.30 K/GPa$  derived for the Hg-based
HTSC. Now , with the values of $\partial \Delta(0)/\partial P$  and  
$dn/dP$ determined, we can calculate
the high order terms of Eq.\ref{coeff} and in particular, we obtain 
$\alpha_2=2.48K/GPa^2$.  The third order coefficient was also calculated
but it does not give any 
appreciable contribution in the range of pressure below $2 GPa$. Thus
with all the coefficients of Eq.\ref{coeff} determined, we are all
set to calculate the values of $T_c(n,P)$.  The calculated  results   
are in excellent agreement with the experimental data\cite{Huang}  as it is
shown in Fig.2. The large value of $dn/dP$ yields a very large
negative second order coefficient $\alpha_2$  what makes the curve change its
derivative at $P \approx 1.2$ GPa. For the Hg compounds the change in 
sign of  $dT_c/dP$ occurs at much larger  $P \approx 30.0$ GPa and from the
above analysis we can see that this difference in behavior is due to the 
large difference
on the value of the  charge transfer $dn/dP$ for this two  compounds. 

\section{Conclusions}

Thus, we  conclude this work pointing out that our novel calculations based on
a BCS type mean field and  on the extended Hubbard Hamiltonian 
with the position of V as a variable parameter, is appropriate 
to describe the interplay between the $T_c$ and hole content for HTSC. 
The change in the position of the attractive interaction  provides
informations on the range of the Cooper pair attractive  mechanism.
Our method yields also a novel  and systematic way to study
the effects of the pressure and gives  indication that
the pressure induces a variation on the attractive potential  
which in turn, gives an intuitive  interpretation on  the
origin of the intrinsic term.  We should also emphasize that our procedure
is {\it based and has its starting point on a microscopic Hamiltonian} and 
it differs from some
pure phenomenological ways to fit the data\cite{Jover} that uses a parabolic
fitting to the experimental pressure data and  are in the
same category of the inverse parabolic fitting for the $T_c \times n$ curves 
mentioned above that are useful
to obtain information on the different sets of data but has not
any microscopic  implications which is  our main concern here.
By applying it to the \Bi superconductors  
and comparing the results  with our previous calculations
on Hg-based HTSC we have seen that the method  not only deals 
successfully with the 
the experimental results but also displays clearly the physical
properties that accounts  for their different experimental 
behavior under pressure. 

\section{Acknowledgments}
We  acknowledge  partial
financial support from the Brazilian agencies Capes and CNPq.

\begin{figure}
{\bf Fig.1} Calculations for the Bi phase diagram.  The continuous line
is the phenomenological fitting and the dashed line are the calculations
described in the article. The
squares are experimental points taken from Ref.13. \\

\noindent{\bf Fig.2} Calculations for the variations of $T_c$ for  the Bi 
compound under pressure. The
circles are experimental points taken from Ref.14.
\end{figure}


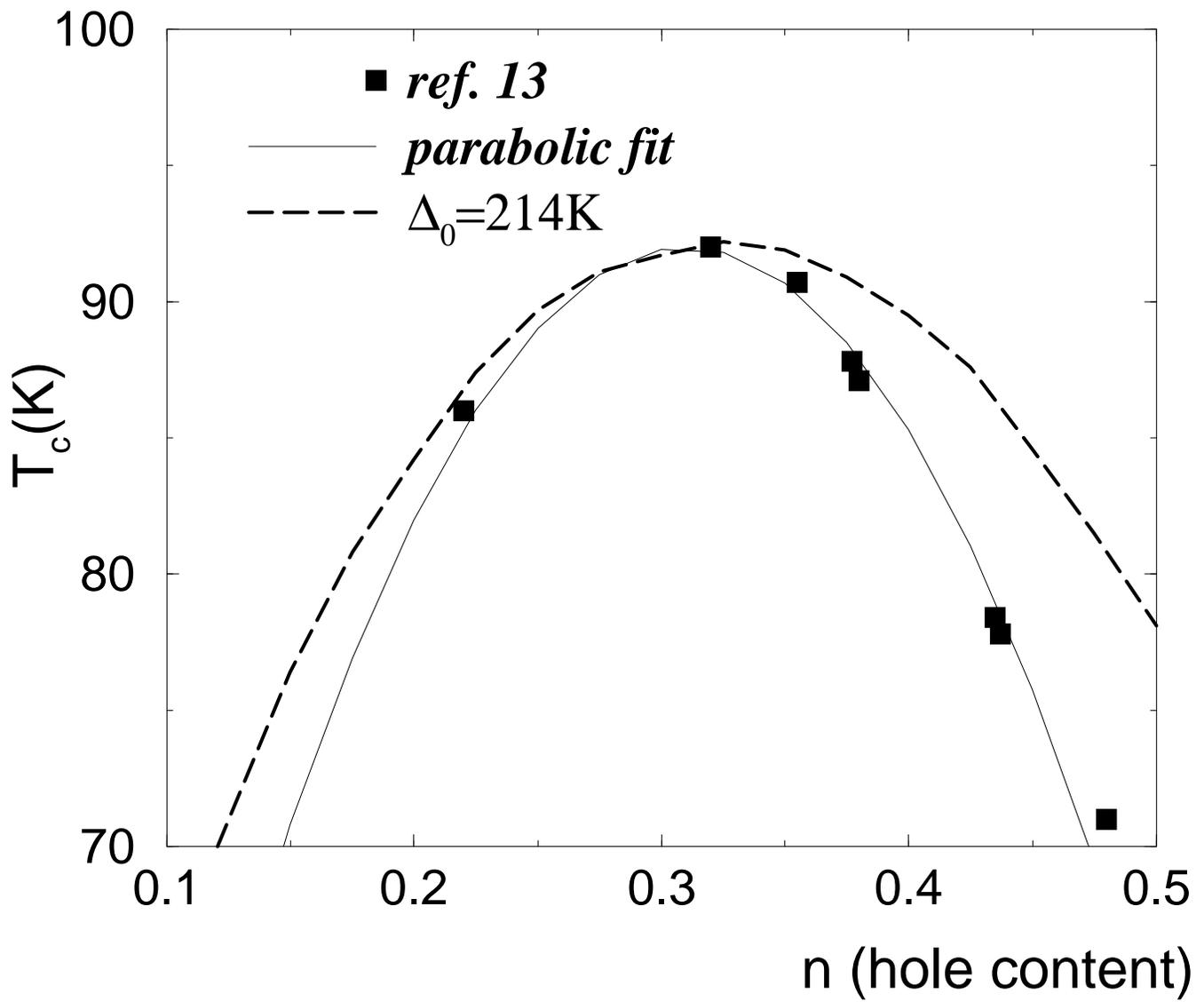
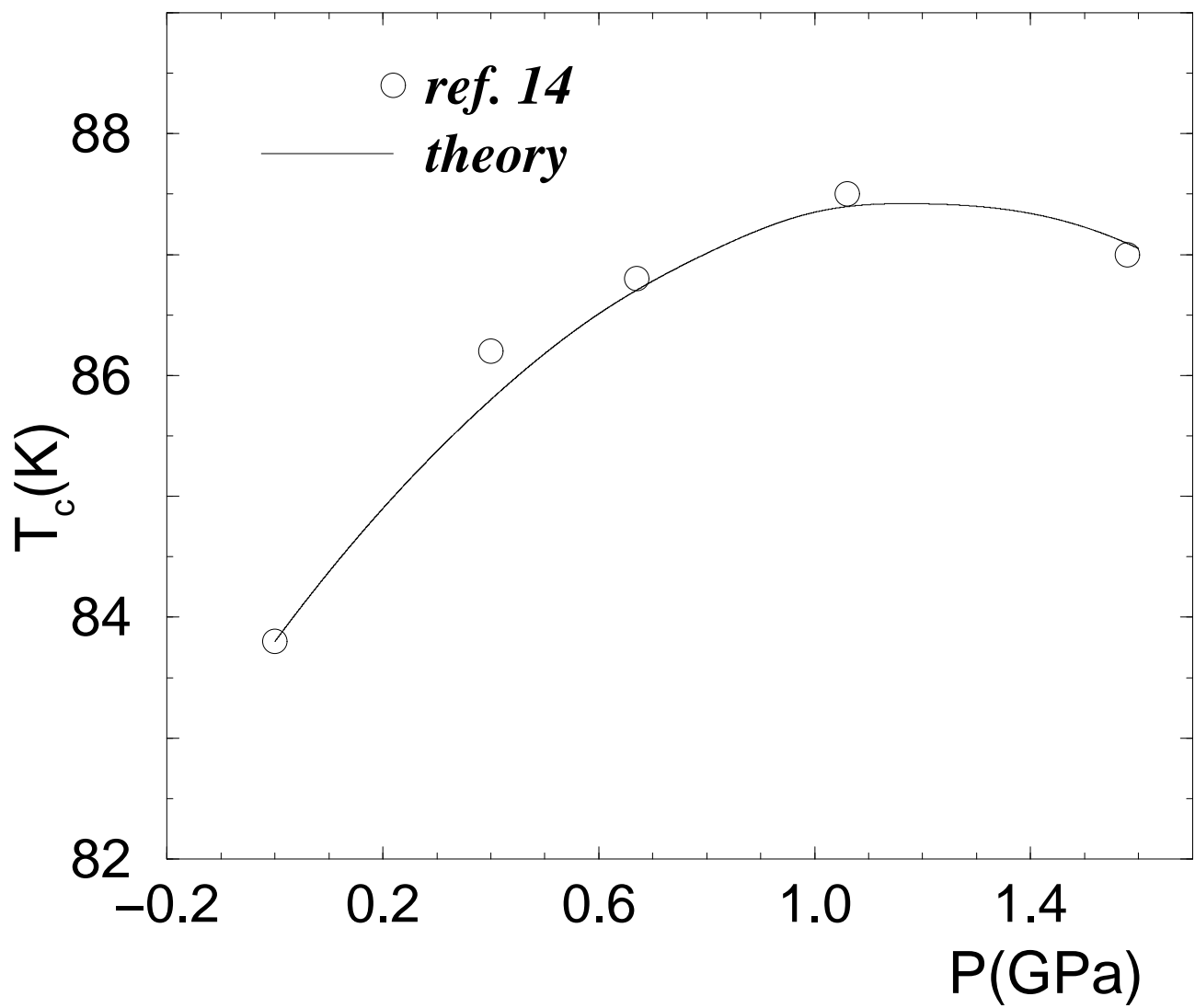
\end{document}